# Growth-sequence-dependent interface magnetism of SrIrO$_3$ - La$_{0.7}$Sr$_{0.3}$MnO$_3$ bilayers

L. Bergmann[1], P. Düring[1], S. Agrestini[2,3], A. Efimenko[4], S. -C. Liao[3], Z. Hu[3], P. Gargiani[2], C. –J. Choi[5], H. Baik[6], D.-S. Park[7], K. Dörr[1], and A. D. Rata[1]

[1]*Institute of Physics, Martin Luther University Halle-Wittenberg, 06099 Halle, Germany*

[2]*ALBA Synchrotron Light Source, E-08290 Cerdanyola del Vallès, Barcelona, Spain*

[3]*Max Planck Institute for Chemical Physics of Solids, Nöthnitzer Strasse 40, 01187 Dresden, Germany*

[4]*European Synchrotron Radiation Facility, Boîte Postale 220, 38043 Grenoble Cedex, France*

[5]*Semiconductor Physics Research Center (SPRC), Chonbuk National University, 567 Baekje-daero, Jeonju, South Korea.*

[6]*Division of Analytical Research, Korea Basic Science Institute, Seoul 136-713, South Korea*

[7]*Laboratory for Electroceramic Thin films, Swiss Federal Institute of Technology–EPFL, Lausanne 1015, Switzerland*

(Abstract)

Bilayers of the oxide 3$d$ ferromagnet La$_{0.7}$Sr$_{0.3}$MnO$_3$ (LSMO) and the 5$d$ paramagnet SrIrO$_3$ (SIO) with large spin-orbit coupling (SOC) have been investigated regarding the impact of interfacial SOC on magnetic order. For the growth sequence of LSMO on SIO, ferromagnetism is strongly altered and large out-of-plane-canted anisotropy associated with lacking magnetic saturation up to 4 T has been observed. Thin bilayer films have been grown coherently in both growth sequences on SrTiO$_3$(001) by pulsed laser deposition and structurally characterized by scanning transmission electron microscopy (STEM) and x-ray diffraction (XRD). Measurements of magnetization and field-dependent Mn L$_{2,3}$ edge x-ray magnetic circular dichroism (XMCD) reveal changes of LSMO magnetic order which are strong in LSMO on SIO and weak in LSMO underneath of SIO. We attribute the impact of the growth sequence to the interfacial lattice structure/symmetry which is known to influence the interfacial magnetic coupling.

I. Introduction

Heterostructures of 3$d$ transition metal oxides (TMOs) with strong electron correlation and Ir-based 5d TMOs, which have a strong spin-orbit coupling (SOC), are increasingly gaining scientific interest [1]. Designing artificial heterostructures of iridates with other well-studied transition metal oxides in order to explore the interfacial coupling between 5d and 3d electrons in oxides is important, both fundamentally and for future applications, since emergent magnetic states are expected. Broken inversion symmetry due to the interface and large SOC result in the occurrence of the Dzyaloshinskii-Moriya (DM) interaction at the interfaces, a powerful effect widely utilized in metallic multilayers and heterostructures to control the magnetic order at interfaces [2, 3]. Joint exchange and DM interactions can induce noncolliner arrangements of spins such as magnetic skyrmions and chiral domain walls associated with topological phenomena [4, 5]. Oxides with noncollinear magnetic order and good electrical conduction are quite rare. Interfaces offer an alternative access to materials with such electronic / magnetic states [6, 7, 8]. Magnetocrystalline anisotropy (MA) is the other fundamental parameter governing the magnetic order and switching fields and is, thus, important for spintronic applications. In oxides, MA at interfaces depends sensitively on lattice structure [9, 10].

Recently, several attempts have been made to modify the electronic structure and magnetism of manganites and ruthenates via coupling with $SrIrO_3$. A topological Hall effect and possible realization of magnetic skyrmions was observed in heterostructures of ferromagnetic $SrRuO_3$ and paramagnetic $SrIrO_3$ and attributed to strong DM interaction at the interface [11, 12]. In heterostructures with ultrathin layers of ferromagnetic $La_{0.7}Sr_{0.3}MnO_3$ and paramagnetic $SrIrO_3$, the magnetic anisotropy was found to undergo a systematic change with the thickness of $SrIrO_3$ [13]. X-ray magnetic circular dichroism (XMCD) measurements revealed the occurrence of a weak interfacial ferromagnetic moment on Ir, which was suggested to be related to the magnetic anisotropy change [13]. Nichols *et al.* found a ferromagnetic ground state in superlattices of (antiferromagnetic) $SrMnO_3$ and $SrIrO_3$, with out-of-plane canting of the magnetic easy axis, and large charge transfer across the interfaces [14]. Recently, a large topological Hall effect has been reported in bilayers of $La_{0.7}Sr_{0.3}MnO_3$ grown on $SrIrO_3$ (2 unit cells) [15]. Mohanta et al. theoretically derive a magnetic phase diagram for $SrIrO_3$ / $La_{0.7}Sr_{0.3}MnO_3$ superlattices which contains spin-spiral and skyrmion phases [16]. Hence, there is increasing insight into the effects of large SOC induced by $SrIrO_3$ on magnetic order of adjacent oxide ferromagnets. It confirms the promise of controllable non-collinear interface spin textures.

We investigate the interfacial coupling between 3d-$La_{0.7}Sr_{0.3}MnO_3$ and 5d-$SrIrO_3$ in bilayer samples with both growth sequences, since earlier work on other oxide interfaces showed clearly different behaviour for the two stacking sequences [6, 17]. The manganite layers of our samples are more than 10 unit cells thick in order to study the interface magnetism of a rather thick layer. Magnetization and field-dependent Mn $L_{2,3}$ edge x-ray magnetic circular dichroism (XMCD) have been measured to characterize the magnetic order. In the bilayers, the soft ferromagnetism of $La_{0.7}Sr_{0.3}MnO_3$ changes towards more hard-magnetic behavior with a canted perpendicular magnetic moment and reduced saturated magnetization. These changes are much stronger in $La_{0.7}Sr_{0.3}MnO_3$ grown on top of $SrIrO_3$ than for the other growth sequence. Based on the observation of similar interface composition by STEM and the structural coherence of the bilayers with the substrate, we conclude on the interfacial lattice structure as origin of the difference between the bilayer types. $SrIrO_3$ is known for its large rotations / tilts of oxygen octahedrons in the orthorhombic structure which may be transferred to the top layer at the interface. Our results demonstrate the importance of the growth sequence for the magnetism of the $La_{0.7}Sr_{0.3}MnO_3$ - $SrIrO_3$ interface with large spin orbit coupling.

**II. Experiment**

Bilayer samples consisting of $La_{0.7}Sr_{0.3}MnO_3$ (LSMO) and $SrIrO_3$ (SIO) layers were grown on $TiO_2$-terminated $SrTiO_3$ (100) substrates (STO) by Pulsed Laser Deposition (PLD) using an excimer laser operating at 248 nm. The growth was monitored *in-situ* by high-pressure Reflection High Energy Electron Diffraction (RHEED) which enables a precise unit cell (uc) control of the layer-by-layer growth. SIO grew in step-flow mode and required a calibration of the growth rate. LSMO layer was grown at 700 °C, while a temperature of 650 °C was used for the growth of SIO layer in an oxygen partial pressure of 0.2 mbar. The laser fluence and repetition rate were 0.8 J/cm$^2$ and 2 Hz, respectively. After growth, samples were annealed for 10 min in oxygen of 200 mbar and cooled to room temperature under same oxygen partial pressure. Bilayers with growth sequences SIO/LSMO/STO(100) and LSMO/SIO/STO(100) and different layer thicknesses have been investigated. The data presented in the figures are from the bilayers used in the XMCD experiment (8 uc of SIO and 12 uc of LSMO). X-ray diffraction (XRD) was employed for *ex situ* investigation of the structural quality and the epitaxy of the films. The XRD measurements were performed with a high resolution Bruker D8 Discover diffractometer using monochromatic Cu $K_{\alpha 1}$ radiation ($\lambda$ = 1.54056 A°). To further analyze the crystalline quality and abruptness of the interfaces, the high-angle annual dark-field (HAADF)-scanning transmission electron microscopy (STEM) images were obtained using a FEI double Cs aberration corrected Titan[3] G2 60-300 S/TEM instrument with Chemi-STEM technology. Magnetization was measured in a Quantum Design SQUID Magnetometer with samples mounted in different orientations to measure in-plane [100] and out-of-plane [001] crystallographic directions. The substrate contribution was subtracted from the field-dependent magnetization data by removing a field-linear diamagnetic term fitted to the high-field data. This procedure also eliminates a part of the paramagnetic SIO magnetization. The XMCD experiments were performed at the BL29 BOREAS beamline at the *ALBA* synchrotron radiation facility. The x-ray absorption was measured using circular polarized light with the photon spin parallel ($\sigma^+$) or antiparallel ($\sigma^-$) with respect to the magnetic field. The spectra were collected with the beam in tilted incidence (20° to the film surface) and in normal incidence. The x-ray beam and the magnetic field were always parallel [18]. The degree of circular polarization delivered by the Apple II-type elliptical undulator was close to 100% for the Mn L-edges. The spectra were recorded using the total electron yield method (by measuring the sample drain current) in a chamber with a vacuum base pressure of $2 \times 10^{-10}$ mbar. The XMCD data were measured at $T$ = 40 K after cooling the samples in a magnetic field of 4 Tesla. The Mn - XMCD hysteresis loops were obtained by measuring, as a function of applied field, the Mn - $L_3$ edge XMCD signal at the energy of the absorption maximum.

### III. Results and discussion
#### A. Structure characterization
##### A1. X-ray diffraction (XRD)

Structural characterization by XRD indicates the high crystallinity and epitaxy. Reciprocal space mapping (RSM) around (-103) Bragg reflection shows that the films are coherently strained on STO(100) substrates, with in-plane lattice parameter of $a$ = 3.905 Å. The RSM of the two bilayer samples (S1: SIO(8uc)/LSMO(12uc)/STO(100) and S2: LSMO(12uc)/SIO(8uc)/STO(100)) with both growth sequences are shown in Fig. 1 A. The vertical alignment of the LSMO, SIO and STO reciprocal lattice point indicates that the bilayers are fully strained on the STO substrates.

### A2. Atomic structure at interfaces (STEM)

Atomic scale characterization was performed by STEM using HAADF-Z contrast. Representative HAADF-STEM images of the samples S1: SIO(8uc)/LSMO(12uc)/STO(100) and S2: LSMO(12uc)/SIO(8uc)/STO(100) are shown in Fig. 1 B. Note that the layers of Ir atoms are very well visible due to the large atomic weight of Ir; the brightness of atomic columns scales approximately with the square of the ordering number of elements. Both bilayers are coherent with the substrate; no lattice defects have been found in STEM images. In the case of sample S1 (Fig.1 B top, left panel), there is a brightness gradient in the LSMO layer which may result from strain or thickness variations in the STEM sample. (Note that an Ir interdiffusion would rather brighten the interface region.) The ferromagnetism of this sample is close to that of single-layer LSMO, indicating little impact of these features on magnetism. At the SIO/LSMO interface, atomic planes of (La,Sr)O-*(Mn,Ir)$O_2$-(Sr,La)O*-Ir$O_2$ are distinguishable (enlarged right panel). Intermixing may occur within the two inner layers, i. e. within 1.5 uc. Beyond these layers, the next two atomic layers on both sides may have a small admixture of Mn in the Ir$O_2$ layer and of Sr in the nominal La$_{0.7}$Sr$_{0.3}$O$_3$ layer. For the bilayer S2 (Fig. 1B bottom, left panel), there is no contrast gradient like in S1 and the sequence of substrate – heavy iridate – lighter LSMO reveals the interfaces clearly. This is the sample for which strong changes of magnetic order are found. The sequence of atomic planes is Ir$O_2$-*(La,Sr)O-(Mn,Ir)$O_2$*-(La,Sr)O at the interface (enlarged right panel). Again, there are two intermixed atomic planes. Intermixing at A sites in the sense of a La:Sr ratio deviating from 70:30 is less easy to identify and could involve the layer marked as (La,Sr)O in the image. We note that the two intermixed interface atomic layers in both samples appear less well defined; this may naturally result from the adaptation of oxygen octahedral rotations / tilts at the interface which are affected by local composition. For this reason, quantitative evaluation of composition with atomic resolution was not possible. Concluding, the chemical composition of LSMO-SIO interfaces for both bilayer sequences appears similar with regard to the metal ions.

## B. Magnetic properties
### B1. Magnetization

Magnetic properties of the same pair of bilayer samples were systematically investigated by measuring the field and temperature dependence of the magnetization with magnetic field along both, the in–plane [100] and out-of-plane [001] directions. Magnetic properties of a reference LSMO single layer grown coherently on STO(001) with a thickness of 12 uc were measured using the same experimental conditions.

Bulk LSMO is a soft ferromagnetic material with high spin polarization at room temperature and a Curie temperature ($T_C$) of 370 K. The magnetization of thin LSMO/STO(100) films lies in the film plane with the easy axis along the [110] direction and the magnetic anisotropy is weak [19]. SIO is a correlated paramagnetic semimetal; SIO thin films do not show long-range magnetic order down to low temperatures [20, 21]. Previous reports demonstrate a small magnetic moment of ~0.05 $\mu_B$/Ir at the interface coupled antiparallel to the Mn moment [13, 14]. Therefore, we expect SIO to contribute very little to the measured total magnetization. In field-dependent measurements, magnetization is attributed to LSMO only.

Fig. 2a,b shows the magnetization loops M(H) for the bilayer samples (S1 and S2) measured along in-plane [100] and out-of-plane [001] directions at 10 K. In Fig.2c we plot for comparison the in-plane magnetization data of a single LSMO reference film together with the in-plane magnetization loops of the S1 and S2 bilayer samples. The LSMO reference film has soft magnetic properties with small coercive field of about 5 mT at 10 K and a saturation magnetization of 463 emu/cm$^3$, similar to other literature reports [19]. The in-plane magnetization of the S1 bilayer is slightly reduced compared to that of the LSMO reference film (see Fig. 2c). Sample S1 has a saturation magnetization of 410 emu/cm$^3$ in a field of 1 T. On the other hand, a drastic reduction of the in-plane magnetization is observed for the S2 bilayer which has a saturation magnetization of 216 emu/cm$^3$ in a magnetic field of 1 T. We note that the saturation in large fields is artificially achieved for sample S2 through the correction procedure for the substrate contribution, see the Experiment section. More insights are obtained from the out-of-plane magnetization loops. Contrary to the in-plane easy axis of a strained LSMO single layer, large out-of-plane remanence is observed for both bilayers (Fig. 2a,b, black curves). For the S1 bilayer, hysteretic behavior with a remanent out-of-plane magnetization of ~30% of the value of 304 emu/cm$^3$ obtained in a field of 1T is found. The S2 sample has a remanence of 37% of the out-of-plane magnetization of 168 emu/cm$^3$ measured

in 1T. This gives evidence for substantial out-of-plane canting of Mn spins in remanence, which is larger for sample S2. Another change is observed in the coercivity ($H_c$) of the bilayer samples. For sample S1, $\mu_0 H_c$ = 47 mT, while in sample S2, $\mu_0 H_c$ is increased to 102 mT at 10 K. The large $H_c$ of LSMO interfaced with SIO is in contrast to the soft magnetic behavior of LSMO single layers ($\mu_0 H_c$ = 5 mT). These results reveal a pronounced change of the magnetic anisotropy of the LSMO layers in both bilayers. The canted out-of-plane spins arise at the LSMO-SIO interface as a consequence of coupling to the SIO layer. The LSMO/SIO/STO(100) sample may even be seen as a 'hard magnet', considering the low magnetization value reached in 1 T at 10 K.

The temperature dependence of the magnetization displayed in Fig. 3 goes along with these observations. The increase of measured magnetization towards low temperatures results, in part, from the paramagnetic contribution of the SIO layer. This SIO contribution cannot be separated, because LSMO is not fully ferromagnetically ordered in the bilayers and, thus, changes its magnetization at low temperatures. In a moderate measuring field of 0.01 T, S1 shows a Curie temperature ($T_C$) of 290 K which is slightly smaller than that of a strained LSMO/STO(001) single film of 12 uc thickness. S2 has a low magnetization in 0.01 T which rises gradually towards low temperatures. The canted out-of-plane magnetization with temperature-dependent magnetic anisotropy is the likely cause of the gradual M(T) rise in sample S2. Hysteretic magnetization curves with remanence are still found at 200 K (not shown).. The M(T) curves show that the anisotropy changes induced in LSMO through the adjacent SIO layer persist up to the Curie temperature.

### B2. X-ray magnetic circular dichroism (XMCD)

To gain more insights into the altered magnetic order of LSMO magnetically coupled to SIO, XMCD measurements were performed at the Mn $L_{2,3}$ edges on the same samples S1 and S2. The XMCD technique has the advantage of tracking the element-specific magnetization of Mn ions as a function of magnetic field without perturbing contributions from the substrate or the SIO layer. Regarding the probed thickness of the bilayers, results from sample S1 with LSMO bottom layer are similar to the behavior of single LSMO; this gives evidence for a sufficient probing depth for the characterization of the bilayer volume. The Mn-XMCD hysteresis curves measured at 40 K are shown in Fig. 4 with increasing and decreasing magnetic field, respectively. For comparison, Fig. 5 presents magnetization loops of the samples recorded at 40 K. Before each XMCD measurement, the sample was cooled in a field of 4 Tesla. The data confirm and reveal more clearly the differences between the two bilayer sample types. In particular, the high-field part up to 4 T provides meaningful information about the saturation which is impossible to evaluate based on SQUID magnetization data. On the other hand, we note that low-field XMCD data (< 0.1 T) are subject to random errors visible in the scattering in Fig.4; hence, coercive fields at 40 K cannot be precisely determined.

For the S1 bilayer, the in-plane XMCD signal exhibits saturated magnetization below 0.5 T (Fig. 4a). The out-of-plane XMCD signal (Fig. 4b) essentially reflects a rotation of Mn moments by the field which is finished near 1.5 T. The remanence in Fig. 4b appears to be very small, which appears in conflict with M(H) data in Fig. 5. However, the coercivity is such low (Fig. 5a, red curve) that XMCD data scattering obscures information about the true remanence. In comparison to LSMO single film XMCD data [22], our S1 data are similar with somewhat larger saturation fields. The coupling to a SIO top layer has moderate impact with regard to increasing the strength of magnetic anisotropy and tilting the easy axis out of the film plane, averaged over a 12 uc thick LSMO film. (Note that the interface is probed with larger sensitivity than the bulk of the LSMO layer, since it is at the top.)

XMCD results of bilayer S2 (Fig.4c,d) are clearly different and demonstrate the strong impact of the SIO-LSMO interface on magnetic order. Most importantly, there is no saturation of the XMCD signal up to 4 T in [100] and [001] directions at 40 K. Magnetic anisotropy of LSMO is sufficiently strong to prevent Mn moments from aligning in field direction in 4 T, reflecting an impressive increase compared to magnetic anisotropy of single LSMO films. Further, the Mn XMCD of sample S2 shows hysteresis with roughly equal remanence along both directions, in-plane [100] and out-of-plane [001], as observed as well in the SQUID magnetization (see Fig.5b). This confirms strong canting of the Mn moments towards the out-of-plane direction, underlining the modified magnetic anisotropy of the 12 uc thick LSMO layer upon interfacial coupling with SIO.

### C. Discussion

Our results reveal strong changes of the magnetic order and anisotropy of thin (12 uc) ferromagnetic LSMO coherently grown on SIO. Structural characterization by x-ray diffraction and STEM shows high film quality, excluding interdiffusion or extended defect formation as source of the altered magnetism. The ordered magnetic moment at low temperatures (10 K, 40

K) is only about half of the LSMO moment for collinear magnetic order (3.7 $\mu_B$ / Mn, 470 emu /cm$^{-3}$). This can be interpreted as very large canted-out-of-plane magnetic anisotropy induced at the interface with SIO. In the light of recent insights on SIO/LSMO interfaces [15, 16], the formation of non-collinear spin structures with large stability in an applied magnetic field is likely to occur. The large out-of-plane canting of magnetic moments is consistent with that. In some distance from the interface, a return to the coherently strained bulk behavior is expected for LSMO, implying a return to collinear ferromagnetic order. This distance is presently unknown, it is possible that the layer thickness of 12 uc is smaller.

For the growth sequence of a SIO top layer on LSMO, all changes of LSMO magnetism are much weaker. True magnetic saturation is reached in 1.5 T (out-of-plane) and 0.5 T (in-plane), indicating collinear magnetic order induced by the applied field. There is a clear difference to the other growth sequence. We conclude that also for this oxide interface, interfacial magnetism depends heavily on the stacking sequence as was reported earlier for LSMO/SrRuO$_3$ [6] and LaMnO$_3$/LaNiO$_3$[17]. This means that superlattices are likely to contain two different interface types. Further, it has consequences for using the SIO interfacing in electronic devices, where large effects like the recently observed large spin-orbit torque, for Permalloy on SIO [23], can be expected for the LSMO layer on top of SIO.

The major result of the present work is experimental finding of the impact of growth sequence. Regarding the origin of this observation, we suggest a physical mechanism which is hypothetical. Due to the large SOC of SIO, the DM interaction influences the magnetic order and anisotropy of LSMO at the interface with SIO. It tends to induce a noncollinear arrangement of spins and, hence, reduces the magnetization and enlarges the magnetic field required for collinear ordering in the field direction. Therefore, a strong DM interaction can cause the observed magnetization characteristics.

The different interface magnetism in dependence on growth sequence must rely on a structural or chemical difference of the interfaces. X-ray and STEM analyses gave no indication of in-plane strain relaxation or differences in chemical composition for the interface types. The latter does not include oxygen; further, the interdiffused two atomic planes could not be analyzed quantitatively. Hence, in-depth structural characterization of SIO-LSMO interfaces is yet to be done. However, there is another structural parameter which is expected to strongly affect DM interaction. At coherent oxide interfaces, rotations and tilts of oxygen octahedrons (OOR) need to elastically adapt. OOR alter the strength of the DM interaction, because the DM interaction depends heavily on the M-O-M bond angle (with metal ion M = Ir or Mn) [24,25]. SIO has large OOR in its orthorhombic phase [23, 26], whereas LSMO has much weaker OOR. For a considerable number of oxide interfaces, the transfer of OOR from the underlayer to the next layer in an interfacial region has been demonstrated [10,27,28,29]. Therefore, we hypothetically suggest that OOR in LSMO at the interface with SIO depend on growth sequence, leading to different strength of the interfacial DM interaction.

## IV. Conclusions

To summarize, bilayer samples consisting of $La_{0.7}Sr_{0.3}MnO_3$ and $SrIrO_3$ layers were grown fully strained on $TiO_2$-terminated $SrTiO_3$ (100) substrates by pulsed laser deposition with both growth sequence. Measurements of magnetization and field-dependent Mn $L_{2,3}$ edge x-ray magnetic circular dichroism reveal strong changes of $La_{0.7}Sr_{0.3}MnO_3$ magnetic order. The soft ferromagnetism of $La_{0.7}Sr_{0.3}MnO_3$ changes towards more hard-magnetic behavior with a canted perpendicular magnetic moment and reduced saturated magnetization. These changes are much stronger in $La_{0.7}Sr_{0.3}MnO_3$ grown on top of $SrIrO_3$ than for the other growth sequence.

Based on the observation of similar interface composition by STEM and the structural coherence of the bilayers with the substrate, we conclude that the interfacial lattice structure is at the origin of the difference between the two bilayer types. In addition, due to the large SOC of $SrIrO_3$, the DM interaction may as well influence the magnetic order and anisotropy of LSMO at the interface with $SrIrO_3$. Our results demonstrate the importance of the growth sequence for the magnetism of the $La_{0.7}Sr_{0.3}MnO_3$ - $SrIrO_3$ interface with large spin orbit coupling.

## Acknowledgements

The research was supported by *Deutsche Forschungsgemeinschaft, SFB 762 Functionality of Oxide Interfaces* (Projects A9, B1). We thank ALBA synchrotron light facility for providing beamtime and technical assistance.

References

[1] L. Hao, D. Meyers, M.P.M. Dean, J. Liu, *Novel spin-orbit coupling driven emergent states in iridates-based heterostructures,* J. Phys. Chem. Solids (2017) (https://doi.org/10.1016/j.jpcs.2017.11.018) .

[2] A. Fert, V. Cros, and J. Sampaio, *Skyrmions on the track*, Nat. Nanotechnol. **8**, 152 (2013).

[3] G. Chen and A. K. Schmid, *Imaging and tailoring the chirality of domain walls in magnetic films*, Adv. Mater. **27**, 5738 (2015).

[4] N. Nagaosa, X. Z. Yu, and Y. Tokura, *Gauge fields in real and momentum spaces in magnets: Monopoles and skyrmions,* Philos. Trans. R. Soc., A **370**, 5806 (2012).

[5] N. Nagosa and Y. Tokura, *Topological properties and dynamics of magnetic skyrmions,* Nat. Nanotechnol. **8,** 899 (2013).

[6] S. Das, A. D. Rata, I. V. Maznichenko, S. Agrestini, E. Pippel, N. Gauquelin, J. Verbeeck, K. Chen, S. M. Valvidares, H. Babu Vasili, J. Herrero-Martin, E. Pellegrin, K. Nenkov, A. Herklotz, A. Ernst, I. Mertig, Z. Hu, and K. Dörr**,** *Low-field switching of noncollinear spin texture at $La_{0.7}Sr_{0.3}MnO_3$-$SrRuO_3$ interfaces*, Phys. Rev. B **99**, 024416 (2019).

[7] J. D. Hoffman, B. J. Kirby, J. Kwon, G. Fabbris, D. Meyers, J.W. Freeland, I. Martin, O. G. Heinonen, P. Steadman, H. Zhou, C.M. Schlepütz, M. P. M. Dean, et al, *Oscillatory Noncollinear Magnetism Induced by Interfacial Charge Transfer in Superlattices Composed of Metallic Oxides*, Phys. Rev. X **6**, 041038 (2016).

[8] B. Li, R. V. Chopdekar, E. Arenholz, A. Mehta, and Y. Takamura, *Unconventional switching behavior in $La_{0.7}Sr_{0.3}MnO_3$/$La_{0.7}Sr_{0.3}CoO_3$ exchange spring bilayers*, Appl. Phys. Lett. **105**, 202401 (2014).

[9] Z. Liao, M. Huijben, Z. Zhong, N. Gauquelin, S. Macke, R. J. Green, S. Van Aert, J. Verbeeck, G. Van Tendeloo, K. Held, G. A. Sawatzky, G. Koster, and G. Rijnders, *Controlled*

*lateral anisotropy in correlated manganite heterostructures by interface-engineered oxygen octahedral coupling*, Nat. Mater. **15**, 425 (2016).

[10] D. Kan, R. Aso, R. Sato, M. Haruta, H. Kurata, and Y. Shimakawa, *Tuning magnetic anisotropy by interfacially engineering the oxygen coordination environment in a transition metal oxide*, Nat. Mater. **15**, 432 (2016).

[11] J. Matsuno, N. Ogawa, K. Yasuda, F. Kagawa, W. Koshibae, N. Nagaosa, Y. Tokura, and M. Kawasaki, *Interface-driven topological Hall effect in $SrRuO_3$-$SrIrO_3$ bilayer*, Sci. Adv. **2**, e1600304 (2016).

[12] Y. Ohuchi, J. Matsuno, N. Ogawa, Y. Kozuka, M. Uchida, Y. Tokura, and M. Kawasaki, *Electric-field control of anomalous and topological Hall effects in oxide bilayer thin films*, Nat. Commun. **9**, 213 (2018).

[13] D. Yi, J. Liu, S.-L. Hsu, L. Zhang, Y. Choi, J.-W. Kim, Z. Chen, J. D. Clarkson, C. R. Serrao, E. Arenholz, P. J. Ryan, H. Xu, R. J. Birgeneau, R. Ramesh, *Atomic-scale control of magnetic anisotropy via novel spin–orbit coupling effect in $La_{2/3}Sr_{1/3}MnO_3$/$SrIrO_3$ superlattices,* Proc. Natl. Acad. Sci. U. S. A. 113 6397 (2016).

[14] J. Nichols, X. Gao, S. Lee, T. L. Meyer, J.W. Freeland, V. Lauter, D. Yi, J. Liu, D. Haskel, J. R. Petrie, E.-J. Guo, A. Herklotz, D. Lee, T. Z. Ward, G. Eres, M. R. Fitzsimmons, H. N. Lee, *Emerging magnetism and anomalous Hall effect in iridate manganite heterostructures*, Nat. Commun. **7**, 12721 (2016).

[15] Yao Li, Lunyong Zhang, Qinghua Zhang, Chen Li, Tieying Yang, Yu Deng, Lin Gu, and Di Wu, *Emergent Topological Hall Effect in $La_{0.7}Sr_{0.3}MnO_3$/$SrIrO_3$ Heterostructures*, *ACS Applied Materials & Interfaces 11* (23), 21268 (2019).

[16] Narayan Mohanta, Elbio Dagotto, and Satoshi Okamoto*, Topological Hall effect and emergent skyrmion crystal at manganite-iridate oxide interfaces,* Phys. Rev. B **100**, 064429 (2019).

[17] M. Gibert, M. Viret, A. Torres-Pardo, C. Piamonteze, P. Zubko, N. Jaouen, J.-M. Tonnerre, A. Mougin, J. Fowlie, S. Catalano, A. Gloter, O. Stéphan, and J.-M. Triscone, *Interfacial Control of Magnetic Properties at LaMnO$_3$/LaNiO$_3$ Interfaces,* Nano Lett. 15, 7355 (2015).

[18] S. Agrestini, Z. Hu, C.-Y. Kuo, M. W. Haverkort, K.-T. Ko, N. Hollmann, Q. Liu, E. Pellegrin, M. Valvidares, J. Herrero-Martin et al.*, Electronic and spin states of SrRuO$_3$ thin films: An x-ray magnetic circular dichroism study,* Phys. Rev. B **91**, 075127 (2015).

[19] H. Boschker, M. Huijben, A. Vailionis, J. Verbeeck, S. Van Aert, M. Luysberg, S. Bals, G. Van Tendeloo, E. P. Houwman, G. Koster, D. H. A. Blank, G. Rijnders**,** *Optimized fabrication of high-quality La$_{0.67}$Sr$_{0.33}$MnO$_3$ thin films considering all essential characteristics*, J. Phys. D: Appl. Phys. **44**, 205001, (2011).

[20] A. Biswas, and Y. H. Jeong, *Growth and engineering of perovskite SrIrO$_3$ thin films*, Curr. Appl. Phys. **17** , 605 (2017).

[21] P. Schütz, D. Di Sante, L. Dudy, J. Gabel, M. Stübinger, M. Kamp, Y. Huang, M. Capone, M.-A. Husanu, V. N. Strocov, G. Sangiovanni, M. Sing, and R. Claessen, *Dimensionality-Driven Metal-Insulator Transition in Spin-Orbit-Coupled SrIrO$_3$,* Phys. Rev. Lett. **119**, 256404 (2017).

[22] D. Pesquera, G. Herranz, A. Barla, E. Pellegrin, F. Bondino, E. Magnano, F. Sanchez, and J. Fontcuberta, *Surface symmetry-breaking and strain effects on orbital occupancy in transition metal perovskite epitaxial films,* Nat. Comm. **3**, 1189 (2012).

[23] T. Nan et al., *Anisotropic spin-orbit torque generation in epitaxial SrIrO$_3$ by symmetry design,* Proc. Natl. Acad. Sci. U. S. A. **116**, 16186 (2019).

[24] S. Dong, K. Yamauchi, S. Yunoki, R. Yu, S. Liang, A. Moreo, J.-M. Liu, S. Picozzi, and E. Dagotto, *Exchange Bias Driven by the Dzyaloshinskii-Moriya Interaction and Ferroelectric Polarization at G-Type Antiferromagnetic Perovskite Interfaces,* Phys. Rev. Lett. **103**,127201 (2009).

[25] S. Dong, Q. Zhang, S. Yunoki, J.-M. Liu, and E. Dagotto, *Ab initio study of the intrinsic exchange bias at the SrRuO3/SrMnO3 interface*, Phys. Rev. B **84**, 224437 (2011).

[26] J. G. Zhao, L. X. Yang, Y. Yu, F. Y. Li, R. C. Yu, Z. Fang, L. C. Chen, and C. Q. Jin, *High-pressure synthesis of orthorhombic SrIrO$_3$ perovskite and its positive magnetoresistance,* J. Appl. Phys. **103**, 103706 (2008).

[27] E. J. Moon, R. Colby, Q. Wang, E. Karapetrova, C. M.Schlepütz, M. R. Fitzsimmons, and S. J. May, *Spatial control of functional properties via octahedral modulations in complex oxide superlattices*, Nat. Commun. **5**, 5710 (2014).

[28] E. J. Moon, P. V. Balachandran, B. J. Kirby, D. J. Keavney, R. J. Sichel-Tissot, C. M. Schlepütz, E. Karapetrova, X. M. Cheng, J. M. Rondinelli, and S. J. May, *Effect of interfacial octahedral behavior in ultrathin manganite films*, Nano Lett. **14**, 2509 (2014).

[29] J. Chakhalian, J. W. Freeland, A. J. Millis, C. Panagopoulos, and J.M. Rondinelli, *Emergent properties in plane view: Strong correlations at oxide interfaces*, Rev. Mod. Phys. **86**, 1189 (2014).

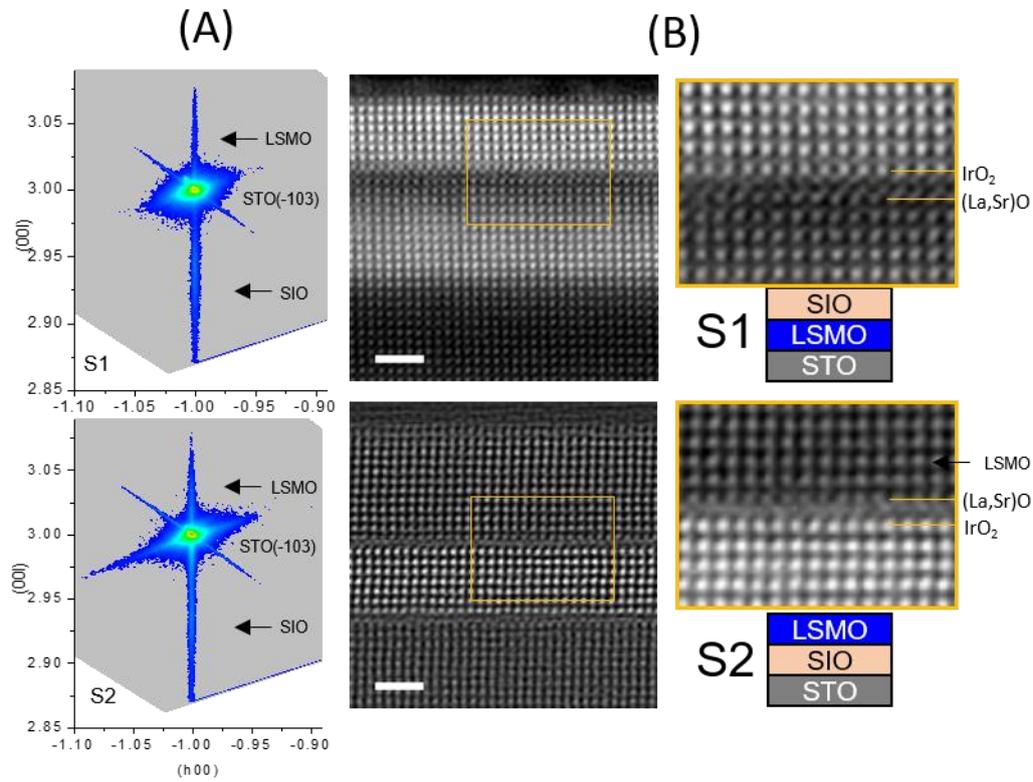

Fig. 1 (A) Reciprocal space mapping (RSM) around (-103) Bragg reflection of the S1: SIO/LSMO/STO(100) and S2: LSMO/SIO/STO(100) bilayers.
(B) High-angle angular dark-field (HAADF) STEM images of the S1: SIO/LSMO/STO(100) and S2: LSMO/SIO/STO(100) bilayers. The high-magnification images correspond to the marked regions of LSMO/SIO interface. Scale bars in the images are 2 nm.

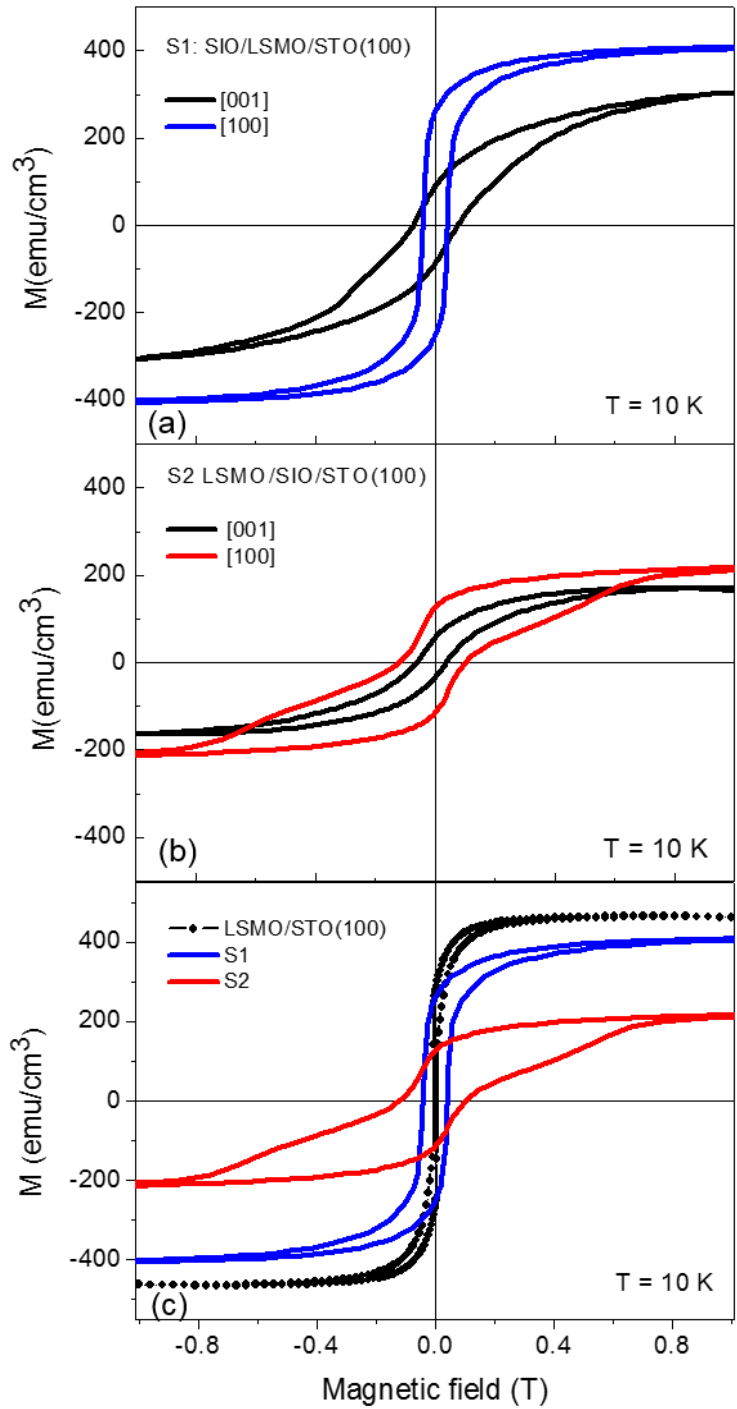

Fig. 2 Magnetic hysteresis loops along the in-plane [100] and out-of-plane [001] directions of (a) S1: SIO/LSMO/STO(100), (b) S2: LSMO/SIO/STO(100) bilayers and (c) in-plane [100] magnetization of S1, S2 and reference single layer LSMO (12 uc) coherently strained on STO(100), measured at 10 K.

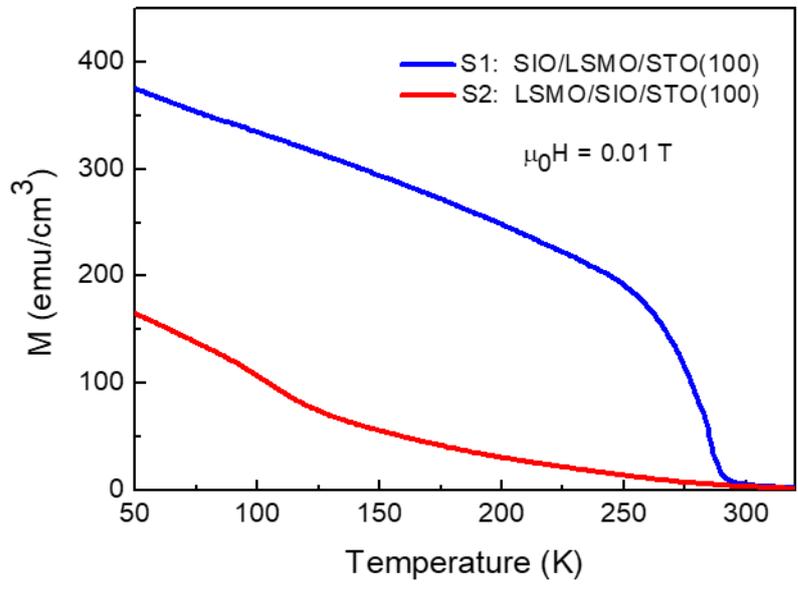

Fig. 3 Temperature dependence of the magnetization of the two bilayer samples S1 and S2 measured during warming in 0.01 T.

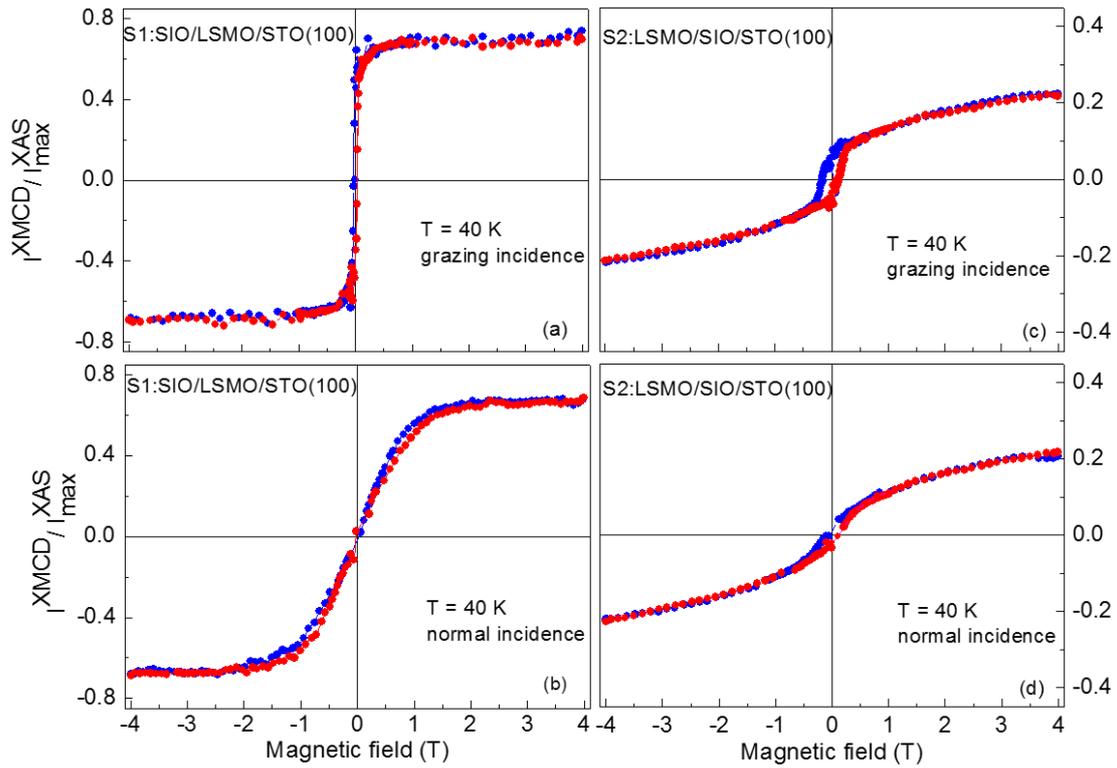

Fig.4 Element-specific Mn $L_{2,3}$ edge magnetic hysteresis loops of the S1 and S2 bilayers measured in grazing incidence (a and c) nad normal incidence (b and d) at 40 K. The magnetic field is swept from +4T to -4T (blue curves ) and from -4T to +4T (red curves).

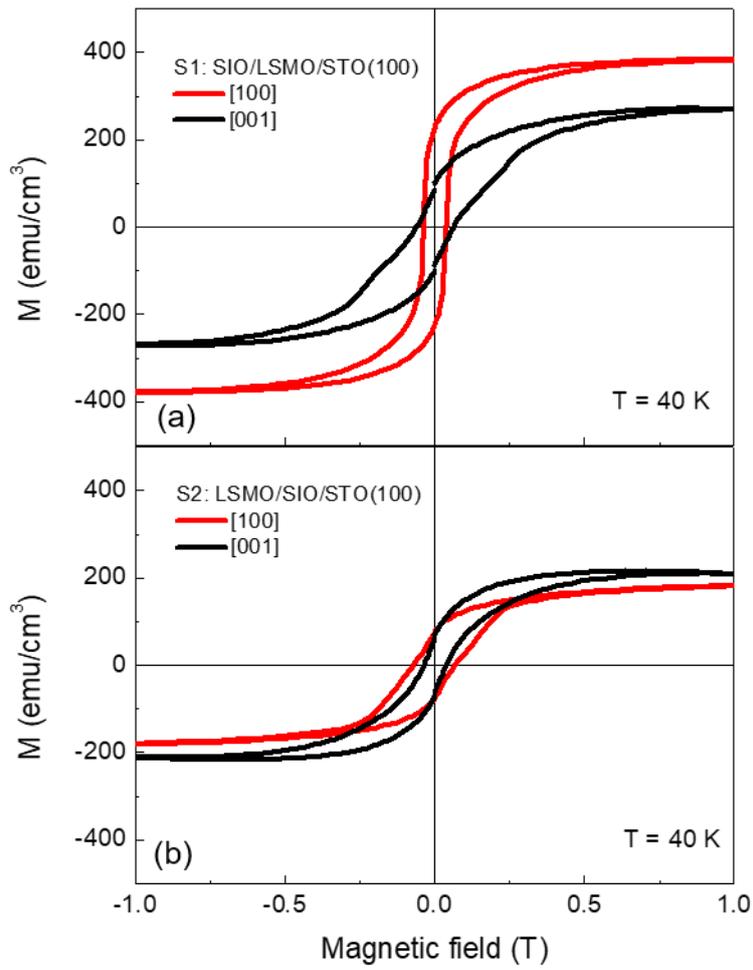

Fig. 5 Magnetic hysteresis loops along the ip-plane [100] and out-of-plane [001] directions of (a) S1: SIO/LSMO/STO(100) and (b) S2: LSMO/SIO/STO(100) bilayers, measured at 40 K.